Consciousness: From the Perspective of the Dynamical Systems Theory

By
Jahan N. Schad, PhD
Retired LBNL (UCB) Scientist
376 Tharp Drive, Moraga, Ca 94556
Email: Jaschadn@gamil.com
Phone: (925) 376-4126

Abstract

Beings, animate or inanimate, are dynamical systems that continuously interact with the (external and /or internal) environment through the physical or physiologic interfaces of their Kantian (representational) realities. And the nature of their reactions is determined by their systems' inner workings. It is from this perspective that this work attempts to address some of the long held philosophical questions; major one among them consciousness, in the context of the physicality of such systems. And to this end, the approach relies upon the appropriate governing mathematical formalisms of system's operations (behavior): For higher beings, the *computational theory of the brain* (Edelman 1987; Kandel 2013), despite lack of details, provides the necessary insights into the likely mathematical processes which must be behind the operations of the system. For inanimate matter, the process is gravely simpler: the responses to environmental (initial and boundary) conditions (inputs) are governed by its field equation (constitutive properties, and constitutive and conservation laws), which render physical changes, which are the *expressions* (outputs) some of which may appear on their interfaces with their external environment. In the former, that is the case of the higher beings, their systems' operations are generally very complex and inevitably would involve brain (computed) solutions of discerned complexities (from sense organ inputs) and streaming downloads of the results (perceptions/conceptions outputs), through the nervous system, to the body physiologic interfaces, for their *expressions*. The latter expressions are animate functions and characteristics such as biological sustenance; maintenance, behavior, thoughts and vocalizations; and the seemingly awareness of sentience, and other associated phenomena, which together define the consciousness, albeit with some reporting shortcoming due to interface display limitations.

Prima facie, the genesis of the consciousness, from the view point of dynamic system theory, -- being simply *expressions* of some of the results of their interaction with the environment -- allows for the generalization of this phenomenon, which is considered only a higher animate peculiarity, to all matter with spatial representation, --animate or otherwise -- granted with vast differences of the nature, and complexities of the related expressions, some of which in humans are referred to as the *"experiences of consciousness."* In such realm, consciousness is fundamental to all matter with objective and subjective aspects to it: the potential to react signifies the *"objective consciousness;"* and the nature of the reaction defines the *"subjective consciousness."* And it is the specificity of the latter, of whatever nature, which separates the animate from the inanimate existence.

Within the laid out framework of the present theory, the big baffling question of philosophy, as well as how and where the human subjective experiences of consciousness happen, *the hard problem* (Chalmers, 1995), find plausible answers: All aspects of human consciousness, are renditions of the results of some of the brain computed events (perceptions/conceptions), -- in response to external and internal stimuli –- by neural mechanisms (Schad, 2016), as *functions* and *expressions*, in different modes, through various physiological body interfaces. In humans, the utterance interface displays two of the major components of consciousness of special interest to this work; the *thought* and the *vision*: they are certain streamed downloads of perceptions, which are *expressed* by this interface (Schad, 2016), mostly inaudibly; though occasionally sounded off, as loud thinking. However, at times, the complexities of the thought and vision (download) contents, – likely involving an extensive Lexicon – render their occurring audible reporting deficient due to vocal instrument filtering. And this inevitable physiologic shortfall (caused by vocal frequency bandwidth limitation),-- the incompleteness of the audible expressions of subjective consciousness-- *recognized as the hard problem*, is very likely sanctioned by the evolutionary processes due to the absence of any survival value. This proposed system theory approach to the understanding of the human sentience and other facets of the brain (mind), follows and complements the (generally accepted) cognitive sciences reductionist (experimentally based) consensus of absence of free will.

Introduction

*"….Perhaps it will take the thinking in a science such as biology, which is of a more general order than the three with which we have been concerned, psychology, medicine, and sociology, to provide the answer all three are seeking"*
Samuel J. Beck and Herman B, Molish (1959)

Today's much advanced state of knowledge owes much to the symbiotic efforts of the fields of philosophy and sciences, which have continued throughout all ages. However the rapid development of natural sciences, which had started since ninetieth century, and that of the advances of the sciences of the brain that had taken roots early in the twentieth century, have been increasingly influencing philosophy; and been of prodigies help in search for answers to its long held big questions. Nonetheless, as it has always been the case with all the frontiers of knowledge, philosophy remains to continue its synthesis of the facts of the mind phenomena to finally trigger the development of relevant scientific theories. As it stands, the philosophical world still remains with all of its big question such as the *nature of reality; mind-body problem; free will*, etc.; and then the most important one, the beholder of them all; the main characteristic of sentience, the *consciousness.* The very phenomenon of consciousness, at least in case of human beings, has been behind whatever sense of life they have, in general, and, in particular, driving the efforts to divulge its very own rendering. And despite the knowledge of sciences and humanities embedded in it collectively, the puzzle of the promoter itself (the consciousness) remains the major challenge that philosophy, and some of its recent daughter sciences, are facing. The perplexed state of the knowledge, in regards to the questions of the mind, is evinced in the opinion polls taken from philosophers across many world institutions, over past few years (Bourget, and Chalmers 2013): the apparent stagnation, has led to examination of philosophy's reasoning structure; by some of its (today's) brilliant and ardent torch bearers (e.g., Chalmers talk 2014). In such evaluations, the lack of progress is being attributed to the weaknesses in the philosophy's arguable premises for the mind; and for how sentience is to be unraveled. Recent Phil paper surveys results are reflected in David Chalmers (2014) statement:"

*There is no collective convergence about truth of the big questions of philosophy such as mind body, Free will, etc., because there are no indisputable premises (axioms, or a more fitting term "postulate") to base the argumentative approach of philosophy on them, in order to come up with their compelling proof."*

However, there is an exception in this finding and that is the fact that opinions on consciousness are seemingly converging: there is consensus in parsing the difficulties of understanding consciousness into *hard* and *easy* problems, -- from the view point of the involved complexities – which could render them more tractable: The first category is ascribed to the subjective phenomena, "which result from physical processes and yet not explainable by them (Chalmers, 1995);" and therefore, the experiences of consciousness (the hard part), such as feelings, emotions, thought, etc., not explainable because of the absence of any functional attributes, have remained an enigma so far. But the functional events of consciousness (the easy problems), are thought to be possibly explainable by cognitive process of the brain (Chalmers, 1995). The latter designation does not by no means imply that the detail of the related mental operation are known, but the prospects are thought to be favorable, and much better understanding of them is likely to happen within this century (Chalmers 2013). The optimism is based on recent progress in sciences that has opened up the possibilities of achieving some insight into the mysteries of the brain operations in general, and of consciousness phenomena in particular. This opportunity is specifically owed to new understandings of brain functions due to the extensive neurosciences research (Kandel, Schwartz and Jessell 2000), on the one hand, and artificial intelligence successes through deployments of the traditional and the neural/neuronal network computers (Turing 1948, McCulloch 1993; Von Neumann 1958; Churchland 1992; Edelman 1987; Arbib 1989), on the other, which together have led to the formulation of the concept of *the computational brain theory*. This brain theory is presently widely accepted (Kandel 2013; Churchland 1992; Wilczek, 2008), and it's standing according to cognitive neuroscientist Jack Gallant (2016) is the following:

*"Brain is a hierarchically parallel distributed network of tightly interconnected areas feeding forward and feeding backward information all over the place and we really have no concept how such a network should compute information."*

The experimentally established computational brain theory has helped to consolidate some of the divided philosophical schools on the side of determinism (works such as Soon et al, 2008 and Fried et al., 2011), though still falling short of explicating how and where the experiences of consciousness, such as vision and thought, occur. The laborious work of Eric B, Baum entitled in*" what is thought,"* published in 2004, exploits the computational brain theory to address the big questions of the mind, with consciousness among them: The following quotes: *"…computer scientists are confident that thought, and for that matter life, arises from execution of a computer program;"* and *"Mind is flow of information, and consciousness is the experience of the information"* capture some of his very valuable insights, expressed in the work. However, this substantial work, relying on speculative evolutionary biologic claims, and computational principles, falls short of providing indisputable arguments towards achieving its very goal: Compatibilism shortcomings and La Mettrie type confusion persists; and "Cogito Ergo Sum", remains even more vague despite the heroic effort!

Different schools of philosophy such as Materialism, Idealism, Dualism, and Panpsychism, have various palatable takes of the consciousness dilemma; and panpsychism (the belief that everything has a mind), holds that consciousness is an intrinsic property of matter. According to Stanford Philosophy Encyclopedia "*the underlying premise in panpsychism,* at *its very microphysical levels, somehow, builds into animate beings' subjective consciousness experiences.*" However, all the ideas still remain speculative at their core! As a way out of the conundrum, Chalmers (1995) suggests a theory of consciousness that takes conscious experience as a fundamental property of the brain; and further claims that "*we might explain familiar consciousness phenomena involving experience in terms of more basic principals involving experience and other entities.*" He asserts that taking experience, the inseparable feature of life, as an axiom, may provide the basis for a general theory of consciousness. To this end, Chalmers' speculative theory relies on (personal) subjective experience data and on the subjects' verbal reports relating to their experience (Chalmers, 1996). However, the thesis as skillfully as it is put forward, aligned partly with panpsychism philosophy, -- the latter unlike its past is being taken more seriously by other schools of philosophy -- similarly suffers from the arguable premises syndrome.

As such, need for a robust theory that can address the mind and all its attributes, still persists -- a philosopher's stone to be found! Present work is an attempt in meeting the challenge by deploying the functional knowledge of the brain -- what facts the cognitive sciences have established so far -- in a radically different context: that of *the all inclusive computational nervous (central and peripheral) system machinery, in the context of dynamic system theory*; it is in such context that the development of a general *inferential theory of the mind*, with emphasis on consciousness, is aimed at: To this end two available works will be heavily drawn upon, namely, Schad (2016) and Schad (2016), where the natures of the brain computer and its dynamics; and how and where perceptions of thought, vision and other facets of mind occur, are theorized. The general framework of the approach has precedence in the field of cognitive psychology, in what is called the *"Embodied brain"* approach (see Kiverstein and Miller, 2015), which only serves to indicate the coming to terms of the "Field" with the role of *the initial and boundary conditions*; *("the skilled organism environment interactions")*, – familiar to applied physic and engineering community -- on the cognition (computational) processes of the brain, implications of which is of importance to various aspects of human psychological mental states.

In summary, the present work which is based on the scientific inferences drawn from the computational functioning of the central nervous system with brain at the helm, explores brain perception processes, on the basis of the nature of the brain computer and its mathematical computations formalism; and how they give rise to the consciousness; and other philosophically contentious sentience related phenomena -- such as free will, nature of reality, and etc., in the context of a new understanding of the brain operations. Specifically, consciousness is reasoned to be the innate characteristic of all systems, animate or otherwise, which in case of higher beings, humans in particular, draw its essence *autonomously* from their brains. And the work well accords -- contrary to what philosopher D.C. Dennett (1996) suggests– with the take of Charles Darwin (1859), reflected in the following statement:

"*Nevertheless, the differences in mind between man and higher animals, as great as it is, certainly is one of degree and not kind.*"

The Theory

*…" I will write about human beings as though I were concerned with lines and plains and solids"*
Baruch Spinoza (1632-1677 A.D.)

"*Life and soul are one, an animating and expansive force present in everything everywhere*"
Anaximenes (585-528 B.C.)

All Beings animate or inanimate are dynamical systems that continuously interact with their (external/internal) environments through physiological or physical interfaces of their systems' (Kantian) realities. And the nature of their reactions (functions and interface expressions, which evolution deemed necessary), are determined by their systems' inner workings. Given the complexities of most systems, clear understanding of most systems' inner working details, is not generally possible. However, the system functional generalities which have already been established can provide the basis for the development of concepts from the perspective of the dynamical system theory; and such is the basis of the work presented here: In case of inanimate matter, the field equations (constitutive properties, and constitutive and conservation laws) generally allow (analytic/numerical) determination of their systems' reactions to the variations of the environmental conditions, and hence the resulting expressions (behavior). Their systems' physical changes (interface displays), that is, the expressions of their varying reactions, indicate the dynamics of their existence. For higher animate beings, emphatically humans, the systems' behaviors (functional operations) are governed by the mathematical formalisms, which must be (inherently) geared in their *computational brains* (Kandel, Schwartz and Jessell 2000), -- based on cognitive neurosciences understanding – though, the details, which would be the key to the development of a fundamental theory of the mind in general, and consciousness in particular, is not still known. And such details, if known, would encompass knowledge of the nature of the brain computer, i.e., what kind of computer it is; and what mathematical formalism underlies its operations -- considering the obvious complexities involved in reaching this end, it is not likely that any solid understanding of the dynamics of the brain computational operations will be established in the foreseeable future. However, as in all challenges sciences face, the immensity of the task, same also in this case, is not a barrier to a first order attempt of conceiving a plausible theory for the brain functional (computational) operations. The present work embarks toward this goal by *the presumption of likely semblance of the brain computer – from ground level operational perspective -- to those of the (brain inspired) scientific neural and neuronal networks (Edelman, 1987; Arbib 1989)* – the similitude principal (Rayleigh, 1915) adds further credence to this extrapolation. The scientific neural and neuronal networks are well known statistical computational methods -- presently augmented by deep learning (supervised or reinforced) processes -- for the development of the advanced levels of artificial intelligence (AI). Grand multidisciplinary projects, such as Machine Intelligence from Cortical Networks (MCrONS), project (David Cox, Harvard University), are efforts to approach creation of human like intelligence. From the underlying presumptions in such endeavors, it further follows that *the essence of the brain neuronal computation (solution) scheme, at its very fundamental levels, could be likened to that of the most basic scheme (implicitly) involved in the computational operations of the scientific neural networks.* In such layout the brain and the rest of the nervous system, are posited to *discern (resolve) any sense stimulating natural phenomena,* -- that it *to (implicitly) algorithmize them in the infinitude of the discrete synaptic nodal domain of the brain* -- and *to solve (mostly by trial-and error) the resulting equations (Shad, 2016)*, to render perceptions, which define some aspects of the mind and the consciousness: This presumed dynamics behind the operations of the central nervous system, would expectedly accord with the following premises:1) the autonomous data processing and computations in the brain, would provide possible *solutions* for various complexities, *discerned* in various states of their manifestations; 2), streaming downloads of the results via the nervous circuitry, would render *expressions* -- at the human interfaces-- of animation; functions such as biological sustenance, maintenance and behavior; and the awareness of sentience, and its associated phenomena, which define the consciousness phenomenon; and 3), expressions of consciousness would be limited due to the inevitable interface (frequency bandpass) filtering of the massive volume of streamed downloads of the results of the brain computations – such drawbacks are normally expected in any input/output system.

Consciousness development relies on the simulations (computations) in the brain, which renders recreations of events, phenomena and the world, which all together make up the experiences of being. The simulations, of whatever nature, are most likely the processing and executions of the life span learned, and evolution hacked, neuronal ciphers (patterns, i.e., software and firmware), prompting potentiation, induced in turn by proper expression of genes, at synapses, -- beholden by some of the known and perhaps (98%) unknown (if they are not rubbish junk according to Brenner, 2013) segments of neural DNA -- rendering the excitatory and inhibitory tasks that produces what *"is not a cause, it is an effect,"* as Dennett (20016) puts it. Some of results (outputs), *which find syntactical expressions in thought,* -- in the (possibly vast) lexicon of its language -- are not necessarily fully available for the efflorescence of *talk* -- as known; some are reportable in speech and loud thinking, and some as in feelings, emotions and other experiences, which are not satisfactorily reportable. Within the context of the system approach, it is mostly in thought and talks that consciousness debuts itself. Since thought is a tangible event on which humans have seemingly some controls over, the questions of where it occurs, perhaps is one that hits closer to home! Given that thought is the result (output) of the brain computations, there must be an interface where its expression happen: only movable body parts (including facial muscles) and vocal cords are the apparent candidate interfaces to which somatic and visceral output (efferent) signals from the brain, reaches -- by means of motor sensory neurons -- displaying the features of consciousness. Such, has long been recognized as explicitly indicated in Confessions (St. Augustine, 397 AD):

*"And that they meant this thing and no other was plain from the motion of their body, the natural language, as it were, of all nations, expressed by the countenance, glances of the eye, gestures of the limbs, and tone of voice, indicating the affections of the mind, as it pursues, possesses, rejects, or shuns."*

The fact that thoughts are not always vocalized (reported), makes it possible to suggest – drawing upon Ockham's razor principal -- a silent muffled mode for the vocal system -- where the computed thought and thinking appear mainly inaudibly (other physiological displays aside), thus introducing a bi-modal utterance system (activated in either mode by a preamble signal), which act as the display medium; vocal mode for language (Chomsky, 2013) and subvocal mode for thought. Putting it succinctly: Utterance (vocal) Interface is the main venue for outputting brain's results of simulations of the real world and some aspects of interactions with it; in audible (referential or verbal) and mostly inaudible (thought) displays. And the fact that thoughts can always be verbalized amid thinking adds enough confidence to the above claim –- the latter can simply be tested. Another, and further, indication of it is the presence of subvocal activity during thinking that apparently behaviorist took note of long ago, and even went as far as to claim the possibility of decoding it: it is anecdotally reported by Will Durant (1991) in a quote from Spinoza:

*"Have not the behaviorist proposed to detect a man's thought by recording those involuntary vibrations in his vocal cord that seem to accompany all thinking."*

Subvocal Laryngeal (electromyography) recording has been deployed in psychiatric patients for clues for behavioral treatments of Hallucination (Green and Kinsbourne, 1998). Of course detecting (decoding) thoughts from subvocal activities is an enormous task involving stochastic/Neuronal, and more, along the line of the recent work by Nishimoto et al., (2011), which is aimed at "*Reconstructing visual experiences from brain activity evoked by natural movies.*" Quoting Gallant (2016):"

*If there is something in emergent working cognitive memory space, potentially it is decodable information.*"

Perhaps supplementing anatomic MRI (diffusion and functional) efforts, along with the very non-smoothed signals (as opposed to those of voxels) from the subvocal activity, should be a boon to semantic extraction that is pursued in decoding research.

Besides thought, vision is the other very important phenomenon of consciousness: Other than attracting questions about its experience, the complexity of its unconscious development, has led to the general collective assumption of it being a fundamental property of Beings who have eyes; also seemingly, the knowledge of the anatomy and physics of vision's physiologic embodiment (Schwarz S.H. 1999), has served as the convincing rational for the assumption. However, vision, as in thought, begets the questions of how does one see what is seen and where it happens? Addressing the Tactile Vision (Bach-y-Rita 2006), and the Mirror Neuron (Keysers, et. al. 2004) phenomena, Schad (2016) theorizes that brain processes for vision sensation, are basically similar those of tactile sensations, except for involvements of more of the brain's neuronal network and constructs (patterns), in the former. And seeing is but an autonomous recitation (reporting) of the sensation, at the utterance interface -- in the absence of such facility other bodily interfaces, as in case of other beings that lacks it, would perform the task. Also additional evidence pointing at the tactile nature of vision is the fMRI activity of parts of visual cortex during Brail reading by blind subjects (Lipman, 2014). Schad (2016) putts it summarily in the following::

*The experience of vision is in reality just inaudible, and occasionally audible, recital (of its semantics) of the likes of reporting in case of massive cutaneous sensing, and apperception of the environment; and, it is the same, in essence, for all, blind and otherwise, which in the former is understandably drastically limited.*

Therefore, Vision thought, speech and what defines consciousness with all its bells and whistles, and everything that relates to the activities of the mind, are brain perceptions, -- being results of execution of brain programs -- which are broadcasted as *expression* on physiologic interfaces.

The reasoning so far, having laid out a possible rational for how computational brain and the rest of the nervous system-- in the context of dynamic system theory – can account for many aspects of the mind in general, and processes of consciousness, in particular, -- of how and where they materialize, and also provides a plausible logic for resolving the hard problem of consciousness; its subjective aspect, as put by Chalmers (1995):

*"A phenomena which is physically based and yet not explainable by it."*

In the context of the present work, the hard problem finds the following simple explanation:

*Subjective consciousness is the thought expressions of the streaming downloads of the constructed brain perceptions (concomitant with memory registrations), which may only be partially reportable (i.e., some not utterable); due to the complexity of the contents, -- perhaps because of the richness of its lexicon -- and the bandpass limitation of the vocal cords, which could filter them.* At much simpler levels, the inability of verbalization in reproducing of some natural sounds one hears is well known in all languages. This vocal reporting shortcoming could have very likely been sanctioned by the evolutionary processes -- perhaps because reporting to other fellow humans of "what is it to be me," or of "the color perceptions," has had no survival value, at least in the eye blink of time since our appearance on this planet.

Despite the complexity of the environmental interactions of animate matter, and the innate simplicity of the in animate matter, the concept of consciousness can be generalized, to both, from the perspective of dynamic system theory in that they all *react* (respond) to the environmental inputs. This common characteristic, this intrinsic property of all objects with spatial representation, may be referred to as the *objective consciousness;* and the nature of the reaction which separates animate from inanimate, designates their *subjective consciousness* < Fig 1>.

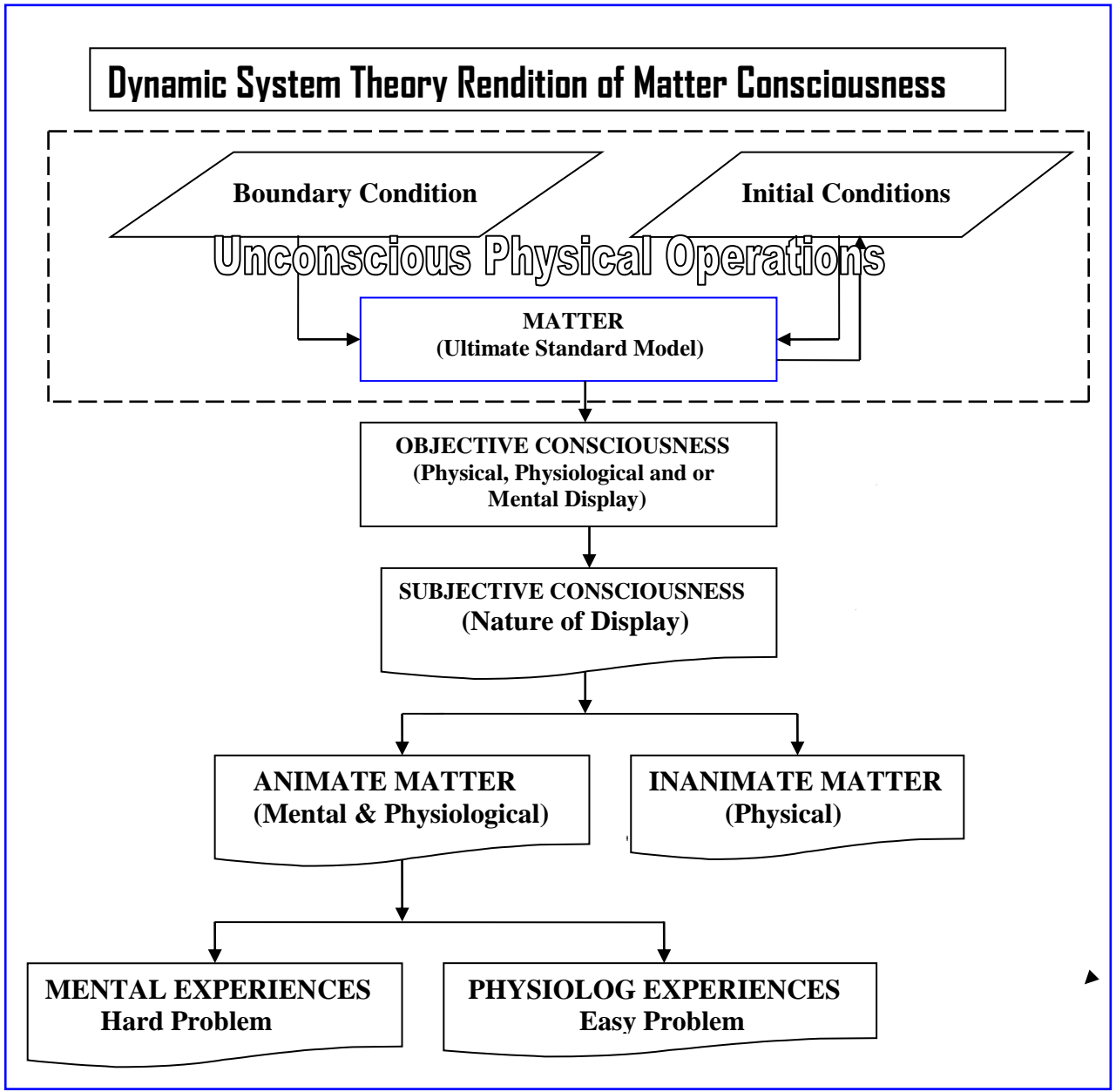

Figure 1- Matter consciousness depiction

This generalized concept of consciousness, accords partly with panpsychism claim; an idea that perhaps has roots in very early thinking: the following quote from Aristotle (Durant, 1991), well speaks to the concept of universality of consciousness, notwithstanding the obvious error:

*"Nature makes so gradual a transition from the inanimate to animate kingdom that the boundary lines which separates them are indistinct and doubtful."*

The generalized consciousness theory is anecdotally evinced in the claims of the experiences of *unison with nature* by subjects in (authority sanctioned) hallucinogenic drug experiments; and by practitioners of intense meditation. In both cases, subjects' brain circuitry gets extensively engaged (chemically affected or overwhelmed), –- evinced by multichannel EEG data (Nader, 2014), and fMRI (NPR's Radiolab report on LSD experiments (2016), and Marina, 1999). And during this process a "hang" situation is very likely occurring in the brain operations: most perception computation come to a halt -- at least in case of meditation with a nonsensical mantra (irresolvable problem), it is likely that the futile brain attempt for solution heavily taxes most of the brain circuitry. Therefore, in such state, output streaming become very limited to a great extent. And of course a time-lapse memory track of the period is registered: perhaps a record of what the universal objective consciousness engenders; the effects of the matter world interaction and perhaps even microphysical entanglement with the environment, when normal sense interactions in the context of physiologic animation are gravely suppressed, or altogether are absent. Obviously, the (ever present unconscious computation operations) circuitry for biological sustenance is not affected during such experiences. Finally, this perspective of consciousness adds much credibility to panpsychism philosophy; perhaps their philosopher's stone is found!

Also this work provides the proper ground for establishing the idea of Man Machine. The concept has its roots in Descartes (1960), and later taken up by La Mettrie, in his "L'Homme Machine (1747)", -- understandably facing insurmountable difficulties -- and shared vigorously by Schopenhauer (1855), and by many (to some degree) in recent years (e.g. Baum (2004); Mlodinow 2011). However, in the context of the present work, the idea proves seemingly very plausible, since it considers higher beings as biological dynamic systems -- with brains (the puppeteer, according to Chomsky, 2016) and the rest of the nervous system, as the control system-- with physiologic mouthpieces, which simply broadcasts their presence. Following quote from the Nobel Laureate Sydney Brenner (Woodham 2014), who in a recorded gathering of scientist, puts the overall claim in the proper context:

*"1) How do the genes specify and build a machine that performs the behavior, and 2) how does the machine perform the behavior? The answer to the first one is we do not know, but the answer to the second one is that it would depend on the queued memory and boundary condition, like any readymade machine."*

The following statement by Philosopher David Hume (1739) sums up the sense of being in the followings:

*"We are nothing but a bundle or collection of perceptions which succeed each other with inconceivable rapidity and are in perpetual flux and movement"*

Experimental Support of the Theory

This proposed approach to the understanding of the human sentience and other facts of life in the context of the dynamic system theory, is partially, though not conclusively, supported by 1), the experimental works of Soon et al, 2008 and Fried et al., 2011, in addressing the experience of will: the latter research summarize its findings, as *"... that the experience of will emerges* as *the culmination of premotor activity (probably in combination with networks in parietal cortex) starting several hundreds of ms before awareness,"* which purports to the underlying essence of the theory; and 2), the result of the analysis of multichannel EEG recordings of subjects during transcendental meditation experiments (Stanford Higher State Lectures), which verifies the reported claims of absence of space and time and body sense, by the coherent Alpha waves; seemingly *a no download episode* in wakeful healthy subjects, while the brain is at full computation capacity, -- resulting from irresolvability of the submitted problem (a nonsensical Mantra) -- causes a "hang" state, when much of the characteristics of sentience disappears.

Conclusion

The computational theory of the brain has been deployed to find answers to some of the long held major questions of philosophy: To this end, three propositions were put forward: 1) that Given the theory, it is the computational outputs of the brain which are relayed through motorsensory neurons to the body's physiological interfaces, which render animation and, in case of many beings, vocalizations, thought, vision, and other effects, together defining the phenomenon of consciousness; and, 2), that, in the case of higher beings, it is the vocal interface, referred to here as utterance interface, which is the *main medium* of expressions of perceptions that broadcasts the conscious mental states in bi-modal, audible and inaudible, modes of the vocal box.; and 3), that *brain is in essence an equation solver*, which discerns being's dynamic environment (as sets of parametric linear equations), through stimulation of body's senses, and solves them (by trial-and-error); and outputs the results as expressions at body's extremities. And that through heredity and learning brain engenders many such equations as readymade patterns (neuronal constructs), available for immediate or fast solutions of discerned problems -- in the likeness of today's deep learning (supervised or reinforced) in Artificial Intelligence developments.

The computational brain theory is further engaged to infer that, 1) despite the complex functional operations of the central nervous system, which render (animate) consciousness (other physiologic activities aside), all matters, animate or inanimate, in the paradigm of the input output systems, have *consciousness;* and 2), that the mere fact of interaction with the environment, defines its *objective* aspect; and what make appearance at the interfaces (the expression), determines its *subjective* aspects, which is a function of the matter's inner workings (physical, or biophysical governing rules); while limited by the medium of its broadcast. In this light, the question of consciousness of all higher beings is also settled.

The imbedded consistency of the approach in the analysis of the nature of consciousness of higher animate matter (much explored by Baum 2004), ), also allows for all aspects of Qualia (regardless of different philosophical takes), as well as providing inroads for all big questions of philosophy. Finally, the reasoning for the concept of universality of consciousness, which also accords with Yogi's claims (based on their repeatable experiences during intensive TM meditation), supports, as well, the main axiom of the Panpsychism theory, and provides philosophy with grounds for unarguable premises.

Acknowledgement

Thanks are due to Dr. A. Hindi, for his enlightening insights and discussions, and to Professor S. Hedayat, for his review of the manuscript and many helpful suggestions.